# DISTANT RING GALAXIES AS EVIDENCE FOR A STEEPLY INCREASING GALAXY INTERACTION RATE WITH REDSHIFT[1]

Russell J. Lavery

Department of Physics and Astronomy, Iowa State University, Ames, IA  50011

Patrick Seitzer

Department of Astronomy, University of Michigan

830 Dennison Bldg., Ann Arbor, MI  48109-1090

Nicholas B. Suntzeff and Alistair R. Walker

NOAO/CTIO, 950 North Cherry Ave., Tucson, AZ  85719

and

G. S. Da Costa

Mt. Stromlo and Siding Spring Observatories, Private Bag,

Weston Creek Post Office, ACT 2611, Australia

ABSTRACT

Hubble Space Telescope WFPC2 images of the Local Group dwarf galaxy Tucana reveal an unbiased sample of distant field galaxies. While a large number of these galaxies can be classified according to standard Hubble types, a significant number of these galaxies can be identified as "ring" galaxies, a morphology likely induced through galaxy collisions. We have identified seven ring galaxy candidates in our fields. Assuming these galaxies lie between the redshifts of 0.1 and 1 and that there has been no evolution in the galaxy interaction rate, then the probability of finding a single ring galaxy in our field of view is less than 1%. Alternatively, if the galaxy interaction rate increases as $(1 + z)^{4.5}$, which represents a high-end estimate of the dependence of the galaxy merger rate on redshift, the probability increases to ~10%. Thus, these observations provide support for a galaxy interaction rate that increases steeply at moderate redshift. We also suggest several additional factors that may play an important role in producing the relatively large number of ring galaxies we have detected.

Subject headings: galaxies: evolution --- galaxies: formation --- galaxies: interactions --- galaxies: peculiar --- galaxies: statistics

1. INTRODUCTION

It is increasingly evident that galaxy interactions and mergers have contributed substantially to the evolution of galaxies, both in terms of their stellar populations and their morphological appearance. The merging of galaxies may have produced many of the elliptical galaxies we observe today in the general field (Toomre & Toomre 1972; Toomre 1977; Schweizer 1983; Schweizer & Seitzer 1992) as well as in rich clusters of galaxies (Lavery & Henry 1988, 1994; Lavery, Pierce & McClure 1992; Couch *et al.* 1994, Dressler *et al.* 1994). While such interactions and mergers were probably much more prevalent in the past, the observational results in support of this supposition have been inconsistent.

The galaxy merger rate as a function of redshift has usually been parameterized in the power law form $(1 + z)^m$. A number of observational programs have been devoted to determining the exponent, $m$, with a fairly wide dispersion in results. Large values of $m$ have been found by Burkey *et al.* (1994), Carlberg (1991), Carlberg, Pritchett & Infante (1994), Yee & Ellingson (1995), and Zepf & Koo (1989), all consistent with a value of $m = 3.5 \pm 1$. Yet, more recently Woods, Fahlman & Richer (1995) find no evidence for an increasing galaxy merger rate. A major difficulty with determining the increasing rate of mergers and interactions with redshift is the difficulty of identifying such systems, even at moderate redshift. Since tidal features and distortions are, in many cases, weak and of low surface brightness, the above programs were based on identifying an excess, compared to local galaxy samples, in the number of galaxy pairs in images of distant field galaxies.

Collisional ring galaxies are a relatively small fraction of all galaxies that have undergone a recent interaction, being produced in the relatively rare circumstance of a small galaxy passing nearly directly through the center of a disk galaxy. The "Cartwheel" galaxy (Zwicky 1941) is probably the most well-known example of a galaxy of this type. A thorough review of the properties of collisionally produced galaxies is given by Appleton & Struck-Marcell (1995). While the ring structure of these galaxies identifies them as galaxies having undergone a recent interaction, at redshifts of $z \sim 0.5$, such galaxies will be only several arcseconds in diameter and thus difficult to identify with even the best image resolution obtainable with ground-based telescopes.

We have obtained deep Hubble Space Telescope WFPC2 images of the Local Group dwarf galaxy in Tucana (Lavery & Mighell 1992) which reveal a large number of stars in this galaxy as well as an unbiased sample of distant field galaxies. The properties of the Tucana galaxy, a dwarf spheroidal classified as dE5 by Lavery & Mighell (1992), will be discussed in Seitzer *et al.* (1996). In this paper, we report finding an unexpectedly large number of ring galaxies in the general field galaxy population lying behind the Tucana dwarf galaxy. The presence of these ring galaxies, having a morphology thought to be the result of galaxy collisions, suggests that collisions between galaxies were much more frequent at earlier epochs and, possibly, that there has also been some evolution in the frequency of various types of interactions.

## 2. OBSERVATIONS

The WFPC2 images of the Tucana Dwarf were obtained during two pointing observations in June 1994 and October 1994. Each data set consisted of four 30 minute F555W ($\sim V$-band) exposures and seven 30 minute F814W ($\sim I$-band) exposures. The second data set was taken at a position angle rotated by 140 degrees from the first data set. This resulted in overlap in the Planetary Camera CCD and in one of the Wide-Field Camera CCDs.

Initial image processing was done through the standard STScI pipeline reduction. The images were combined using a median filter inside the IRAF environment. We adopted the standard photometric zero-points and color coefficients determined by Holtzman *et al.* (1995).

While there are many stars of the Tucana dwarf in the HST/WFPC2 images, many distant background galaxies are easily identifiable. Most can be classified according to standard Hubble types (Griffiths *et al.* 1994; Driver *et al.* 1995). But, while viewing the images during the processing stages, the presence of several ring galaxies was noticed. A thorough visual inspection was then made by one of us (RJL) of the resultant F814W images of each CCD in each data set. A total of seven ring galaxy candidates (RGCs) were identified. These seven galaxies are shown in Fig. 1 with their photometric properties given in Table 1. These magnitudes should be accurate to ±0.1 mag. Tucana is a low luminosity dwarf spheroidal and is relatively metal-poor in composition, with [Fe/H] = −1.8 ± 0.3 (Da Costa 1994; Lavery *et al.* 1996; Saviane, Held & Piotto 1996). Hence, the photometry of the ring galaxies should not be affected by dust in Tucana. In fact, a number of galaxies can be seen through the region of highest stellar density. Of our RGC sample, all but Ring 7 are well away from the core region of Tucana. We identified several additional galaxies which also had ring-like features, but these galaxies showed signs of spiral structure and, therefore, were not included in our ring galaxy sample. The angular diameters of the ring galaxy candidates range from 0.6 to 3 arcsec and we believe our search is complete for ring galaxies having diameters greater than 0.8 arcsec, though this does depend on the inclinations of the galaxies. The angular diameters of the rings and their ($V-I$) colors show a weak correlation with the bluer galaxies being smaller in diameter. Appleton & Marston (1996) have found a similar result for nearby ring galaxies, that those of larger linear diameter are redder in color. But, our correlation of these properties probably results largely from differences in the redshifts for our ring galaxy sample.

Looking at Fig. 1, some of these rings are "empty" while others have nuclei, with Ring 2 being similar in appearance to NGC 985, for example. The two "empty" rings are, not surprisingly, also the bluest galaxies in the sample. Of these seven galaxies, Ring 4 and possibly Ring 1, appear to be O-type rings (see discussion) and may not be the result of an interaction (Few & Madore 1986). Yet, both of these galaxies do have features indicative of a recent tidal interaction.

3. DISCUSSION

The presence of seven RGCs in our relatively small field of view is quite surprising given the rarity of ring systems. To test the significance of this result, we have estimated the likelihood of finding a single ring galaxy in our fields for two scenarios: 1) a galaxy interaction rate that is constant with redshift (the non-evolving case) and 2) a galaxy interaction rate increasing as a function of redshift (the evolving case).

The differential volume element as a function of redshift is given by

$$dV = \frac{c^3 \{q_0 z + (q_0 - 1)[(1 + 2q_0 z)^{1/2} - 1]\}^2}{H_0^3 (1 + z)^3 \, q_0^4 (1 + 2q_0 z)^{1/2}} \, d\Omega \, dz.$$

We have assumed a value of $q_0 = 0.05$ which results in the largest volume along the line-of-sight. Due to both the angular sizes and the magnitudes of these ring galaxies, we have assumed these galaxies to lie between the redshifts of 0.1 and 1 and have integrated the volume element equation through this redshift range.

As the two data sets were taken several months apart, the images were taken at different orientation angles. At these two positions, the region of the sky imaged in the PC field was the same and there was significant overlap between one of the WF CCD fields in each of the data sets. Therefore, our total field of view is equivalent in area to five Wide Field camera CCD fields plus the Planetary Camera CCD field. This is a total angular area of ~5.5 sq. arcmin and is equal to a solid angle, $d\Omega$, of $4.65 \times 10^{-7}$ str.

We have adopted the ring galaxy volume density determined by Few & Madore (1986) of $5.4 \times 10^{-6} \, h^{-3} \, \mathrm{Mpc}^{-3}$, where $h = H_0/100$ km s$^{-1}$ Mpc$^{-1}$. This volume density includes two types of ring galaxies: P-type rings, which have a crisp knotty structure and displaced nucleus, and O-type rings, which have a smooth structure and centrally located nucleus. In the Few & Madore sample, 60% of the ring galaxies were classified P-type while 40% were O-type. The statistical arguments presented by Few & Madore support a collisional origin for the P-type rings. The Few & Madore volume density is consistent with the previous, but less rigorous, determinations of Freeman & de Vaucouleurs (1974) and Thompson (1977).

Substituting these values, we find that for the constant (non-evolving) volume density case the probability of detecting a single ring galaxy in our field is only 0.0075 (<1%). In the second case, we have assumed the ring galaxy volume density increases with redshift as $(1 + z)^{4.5}$. The value of 4.5 for this exponent is one of the higher values for the estimated increase in the galaxy merger rate (Carlberg 1991; Burkey *et al.* 1994; Yee & Ellingson 1995; Zepf & Koo 1989; Carlberg *et al.* 1994). These determinations are claimed to be valid out to a redshift of $z = 0.5$, but we will assume this redshift relation to hold to $z = 1$. The use of this redshift dependence also assumes that the galaxy interaction rate is proportional to the galaxy merger rate and that there has been no change in the frequency of the various types of interactions as a function of look-back time. We find that even including this steeply increasing volume density, the probability of a single ring galaxy being in our HST field is only 0.1 (10%).

We have seven candidate ring galaxies, though it is not likely that all of these are collisionally produced (Ring 4 is quite likely an O-type ring) or independent (Ring 2 and Ring 3 are roughly six ring diameters apart and could have resulted from a single encounter). But, we feel there is little doubt that the majority of these are true ring galaxies. Therefore, we have compared the expectation values determined above to the probability of observing four ring galaxies in our field.

Using Poisson's Law to determine the likelihood of detecting four ring galaxies when the expectation values are 0.01 (the non-evolving volume density case) and 0.1 (the evolving volume density case), we find the probabilities to be $4 \times 10^{-10}$ and $3.8 \times 10^{-6}$, respectively. The extremely low probability for the non-evolving volume density case strongly suggests that the volume density of ring galaxies does increase quite steeply with redshift. But, even including this steep increase of the interaction rate with redshift, the probability of having detected four ring galaxies is still quite small. While the possibility of having observed a 4.6 $\sigma$ region cannot be ruled out, as this is an *a posteriori* statistical argument, there may be several other factors contributing to this small probability:

1) Our sample of rings may be contaminated by misclassified late-type galaxies. This is probably the most major concern given the small angular size of these ring galaxy candidates and the fact that a far-UV (1500 to 2000 Å) image of a galaxy may look quite different from the rest-frame V-band image of the same galaxy. Since the classification of these galaxies was done on the F814W images, this would only be a major concern if the sample of galaxies were at $z \sim 2$ or greater. Our ring galaxies are quite similar in their overall appearance in both filters, suggesting that we are not seeing these galaxies in the far-UV rest-frame wavelength region. As mentioned previously, there were additional galaxies found in the identification process which also had a ring-like appearance. These galaxies were of similar magnitude to our present ring galaxy sample, but were excluded from our sample as there were indications of spiral structure in them. So, while misclassification can be a problem, we have tried to be conservative in the identification of these ring galaxies.

One of the main features indicative of P-type (collisional) rings is the presence of an offset nucleus. For Rings 2 and 6, an offset nucleus is quite obvious, adding credence to the identification of these galaxies as collisional rings.

2) Our sample of rings may be contaminated by planetary nebulae in the Tucana dwarf. We have estimated this possibility by looking at the number of known PNe in other dwarf systems. The stellar population of the Fornax dwarf spheroidal galaxy is similar to that in Tucana and in Fornax, there is one known PN (Danziger *et al.* 1978). The absolute magnitude, $M_V$, of Fornax is approximately −13.5 while for Tucana, $M_V \approx -10$, a factor of 25 in luminosity, suggesting only a 1 in 25 chance (4%) of there being a PN in Tucana. Considering the galactic globular cluster system, there are two known PNe; one in M15 and one in M22. The total luminosity, $M_V$, of the globular cluster system of the Milky Way is about −13. This suggests only a 1 in 8 chance of a PN being present in Tucana. Given these relatively small probabilities, it is unlikely that PNe are a source of contamination in our ring sample.

3) Our assumed redshift interval ($0.1 \leq z \leq 1$) may be too conservative. But, if we assume a larger redshift range for our sample, say out to $z = 2$, there is still only a 3% chance of finding a single ring galaxy in the non-evolving case. For the evolving case with $m = 4.5$, the expected number is actually two galaxies. While this number of galaxies is consistent with our observations, we would caution against accepting this high redshift limit for two reasons. First, the present determinations of $m$ are believed to be valid for redshifts out to $z \sim 0.5$ and assuming this exponent to be valid out to $z \sim 2$ is a rather large extrapolation. Second, the few known examples of normal galaxies at $z \sim 2$ are very compact in appearance with scale sizes of order 0.1 arcsec (Dressler *et al.* 1993; Pascarelle *et al.* 1996), which is not the case for these candidate ring galaxies.

4) In addition to a steeply increasing galaxy interaction rate with redshift, the frequency of different types of galaxy interactions may also change with redshift. Interactions of low angular momentum may have been more frequent in the past, thereby producing relatively more of the plunging encounters necessary for the production of ring galaxies.

5) The galaxy merger rate and the galaxy interaction rate, which we have assumed to follow the same power law, may have different exponents. If the interaction rate increases more steeply than the merger rate, then we would expect the number of nearly direct galaxy collisions to also increase. Presently, we have no means to address this possibility.

While we feel confident that contamination of the RGC sample is not a major concern (points 1 and 2), it is difficult to evaluate which of the other factors is most important in producing the high ring galaxy density we have observed. Given the size of our RGC sample, it is not possible to place any strong constraints on the exponent, $m$, of the galaxy merger rate. However, since most ring galaxies are produced through direct interactions between galaxies, a large sample of distant ring galaxies with measured redshifts would provide an excellent means to determine how the interaction and merger rate has changed with redshift.

## 4. CONCLUSIONS

HST WFPC2 observations of the Tucana Dwarf galaxy have revealed an unbiased sample of moderate redshift field galaxies in which there is a surprisingly large number of ring galaxy candidates, most likely produced by nearly direct galaxy collisions. The presence of these ring galaxies in a random field provides strong evidence for a galaxy interaction rate that increases steeply with redshift, at least as steeply as the average of the determinations for the galaxy merger rate, $(1 + z)^m$ with m = 3.5 ± 1. Other factors, such as a galaxy interaction rate that increases more steeply than the galaxy merger rate or an increase in the fraction of radial, low angular momentum galaxy collisions, may also contribute in producing the relatively large number of observed ring galaxy candidates.

RJL thanks P. Appleton and C. Struck for several productive discussions concerning ring galaxies at high redshift. Support for this work was provided by NASA through grant number GO-5423.02-93A from the Space Telescope Science Institute, which is operated by AURA, Inc., under NASA contract NASS-26555.

TABLE 1

Photometric Properties of the Ring Galaxies

| Galaxy ID | Diameter (") | $V$ | $I$ | $(V-I)$ |
|---|---|---|---|---|
| Ring 1 | 1.5 | 23.78 | 22.00 | 1.78 |
| Ring 2 | 1.0 | 23.62 | 22.22 | 1.40 |
| Ring 3 | 1.1 | 24.14 | 23.35 | 0.79 |
| Ring 4 | 2.9 | 23.02 | 20.53 | 2.49 |
| Ring 5 | 0.8 | 23.88 | 22.55 | 1.33 |
| Ring 6 | 0.8 | 23.25 | 21.81 | 1.44 |
| Ring 7 | 0.6 | 24.51 | 23.54 | 0.97 |

FIGURE CAPTIONS

Fig 1.--The seven ring galaxy candidates identified in the HST WFPC2 images of the Local Group dwarf galaxy Tucana. Each row consists of a low contrast F814W image, a high contrast F814W image, and the F555W image of each galaxy. Each subimage is 6" on a side.